\documentclass{smrauth}
\usepackage{graphicx}
\usepackage{caption}
\usepackage{amsfonts}
\usepackage{amsmath}
\usepackage{latexsym}
\usepackage{multirow}
\usepackage[tight,normalsize,sf,SF]{subfigure}
\begin{document}

\SMR{0}{0}{0}{0}{0}

\title{}{An Analysis of Bug Distribution in Object Oriented Systems}

\author{Alessandro~Murgia\footnotemark[1], Giulio~Concas\footnotemark[2], Michele~Marchesi\footnotemark[3], Roberto~Tonelli\footnotemark[4] and Ivana~Turnu\footnotemark[5].}

\longaddress{Department of Electrical and Electronic Engineering,
University of Cagliari, piazza d'Armi, 09123 Cagliari, Italy.}

\footnotetext[1]{E-mail: alessandro.murgia@diee.unica.it}
\footnotetext[2]{E-mail: concas@diee.unica.it}
\footnotetext[3]{E-mail: michele@diee.unica.it}
\footnotetext[4]{E-mail: roberto.tonelli@diee.unica.it}
\footnotetext[5]{E-mail: ivana.turnu@diee.unica.it}

\received{} \revised{} \noaccepted{}

\begin{abstract}
We introduced a new approach to describe Java software as graph,
where nodes represent a Java file - called compilation unit (CU) -
and an edges represent a relations between them. The software system
is characterized by the degree distribution of the graph properties,
like in-or-out links, as well as by the distribution of Chidamber
and Kemerer metrics computed on its CUs. Every CU can be related to
one or more bugs during its life. We find a relationship among the
software system and the bugs hitting its nodes. We found that the
distribution of some metrics, and the number of bugs per CU, exhibit
a power-law behavior in their tails, as well as the number of CUs
influenced by a specific bug. We examine the evolution of software
metrics across different releases to understand how relationships
among CUs metrics and CUs faultness change with time.\end{abstract}


\section{INTRODUCTION}
Large software systems can be analysed as graphs so huge and intricate
that can be studied using complex network theory.\\
In the case of object oriented (OO) software systems nodes are
the classes or the interfaces, and oriented edges are the
various kinds of relationships between them, inheritance,
composition, dependence. For OO systems there exist also some
consolidated software metrics, also associated to the graph,
usually computed at class level, the most used being the Chidamber and Kemerer (CK) suite of metrics
\cite{Chidamber}. The relationship between metrics and software
quality is fuzzy, and is still the subject of ongoing research.

Related to software quality are software bugs.
Several researchers analysed software evolution in order to
understand the relationship between software management and bug
issues. Purushothaman et al. \cite{Purushothaman} analyzed software
development process to identify what are the relationships between
small changes to the code and bug growth. Kim et al. \cite{Sunghun}
analyzed micro-pattern evolution in Java classes to identify which
of them is more bug-prone. \'{S}liwerski et al. \cite{Sliwerski}
analyzed the fix-inducing changes, i.e. software updates that
trigger the appearance of bugs. In their work, the revision history
associated to compilation units (CUs) was examined to understand
where bugs issues are introduced during CU evolution. Compilation
units, the basic blocks examined in this paper, are files containing
one or more classes, for which it is possible to compute software
metrics similar to those used for classes.

A complete analysis of the relationships
between graph properties of large software systems, statistic of
software metrics, and the introduction and distribution of bugs in
such graphs is, to our knowledge, completely missing.
Zimmerman et al. considered a network analysis on dependences graphs, built on
binary files \cite{Zimmermann}, and how dependencies correlate with,
and predict, defects. Andersson et al. \cite{Andersson} discussed
the Pareto distribution of bugs in classes, without entering into
the details of the statistical properties of software which
determine such distribution. Zhang found that the bug distribution
across compilation packages in Eclipse Java system seems to follow a
Weibull distribution \cite{Zhang}.\\

The aim of this paper is study OO systems using complex
network theory, to improve the knowledge of bugs causes and to
statistically determine their distribution into the system. We extend
the definitions of CK software metrics to CUs to understand the
evolution of faultness, i.e. how a metric variation affects the
number of bugs hitting a CU. A deeper understanding of the dynamics
of software development could be useful for software engineers to
identify which system components will be more prone to
bugs, thus focusing testing and code reviews on these components.\\
We also study the time evolution of software systems and of the
related graphs and metrics, analysing both the source code and the
bugs of various releases of two large Java systems, Eclipse
\cite{Eclipse} and Netbeans \cite{Netbeans}. For each release we
computed the associated software graph and the CK metrics for each
class. Furthermore, we study the number of defects associated to
CUs, as found in the bug-tracking system used for development.

We computed the correlation between
OO metrics and bugs and analyzed the evolution of these
metrics between one release and the next, correlating metrics
changes with the number of defects. We present a scheme of
classification of CUs into categories which allows us to identify
which parts of the software are the most fault-prone, and how these
are correlated to CK software metrics. We
support our findings with significance tests. \\

\section{Method}\label{Method}
We analyze the source code of object-oriented systems written in
Java. Both use CVS as version control system. Eclipse uses Bugzilla
as issue tracker system, while Netbeans uses Issuezilla. The CVS
keeps track of the source code history, Bugzilla and
Issuezilla keep track of the bugs history.
\subsection{Software graph and OO metrics}

An oriented graph is associated to OO software systems,
where the nodes are the classes and the
interfaces, and the edges are the relationships
between classes, namely inheritance, composition and dependence.\\
The number and orientation of edges allow to study the coupling
between nodes. In this graph the in-degree of a class is the number
of edges directed toward the class, and measures how much this class
is used by other classes of the system. The out-degree of a class is
the number of edges leaving the class, and represents the level of
usage the class makes of other classes in the system. In this
context CK suite is a common metrics employed in classes analysis.
We calculated for each node the values of the
four most relevant CK metrics of the associated class:

\begin{itemize}
    \item Weighted Methods per Class (WMC). A weighted sum of all the methods defined in a class. We set  the weighting factor to one to simplify our analysis.
    \item Coupling Between Objects (CBO). The counting of the number of classes which a given class is coupled to.
    \item Response For a Class (RFC). The sum of the number of methods defined in the class, and the cardinality of the set of methods called by them and belonging to external classes.
    \item Lack of Cohesion of Methods (LCOM). The difference between the number of non cohesive method pairs and the number of cohesive pairs.
\end{itemize}

We also computed the lines of code of the class
(LOC), excluding blanks and comment lines. This is useful to keep
track of CU dimension because it is known that a "long" class is
more difficult to menage than a short class.

Every system class resides inside a Java file, called CU. While most
files include just one class, there are files including more than
one class. In Eclipse 10\% of CUs host more than one
class, whereas in Netbeans this percentage is 30\%.
In commit messages issues and issue
fixing always refer to CUs. To make consistent issue tracking with
source code, we decided to extend CK metrics from classes to CUs.
CUs represent therefore the main element of our study. So, we
defined a CU graph whose nodes are the CUs of the system.
Two nodes are connected with a directed edge if at least one
class inside the CU associated with the first node has a dependency
relationship with one class inside the CU associated with the second
node. We refer to this graph for computing in-links and out-links of
a CU-node. We reinterpreted CK metrics onto this CU-graph:
\begin{itemize}
  \item CU LOCS is the sum of the LOCS of classes contained in the CU;
  \item CU CBO is the number of out-links of each node, excluding those representing inheritance. This definition is consistent with that of CBO metrics for classes;
  \item CU LCOM and CU WMC are the sum of LCOM and WMC metrics of the classes contained in the CU,respectively;
  \item CU RFC is the sum of weighted out-links of each node, each out-link being multiplied by the number of specific distinct relationships between classes belonging to the CUs connected to the related edge.
\end{itemize}
For each CU we have thus a set of 6 metrics: In-links, Out-links,
CU-LOCS, CU-LCOM, CU-WMC, CU-RFC and CU-CBO. This was made for all
versions of Eclipse and Netbeans.

\subsection{Bug extraction and metric}
Onto the CU graph we look for nodes hit by Issues. To obtain
this information it is necessary to check the CVS log file, and the
data contained in the ITS.\\
We consider a CU as affected by an Issue when it is modified for
issue fixing. Developers record on the CVS log all fixing activities.
All commit operations are tracked in the CVS log as single entries. Each entry
contains various data, among which the date, the developer who made
the changes, an annotation referring to the reasons of the commit,
and the list of CUs interested by the commit. In case of commits
associated to an issue fixing activity, this is written in the
annotation, though not in a standardized way. It is not simple to
obtain a correct mapping between issue(s) and the related CU(s) \cite{Sliwerski} \cite{Fischer}.

In our approach, we first analyzed
the CVS log, to locate commit messages associated to fixing
activities. Then, the extracted data are matched with information
found in the ITS.
Each issue is identified by a whole
positive number (ID). In commit messages it can appear a string such
as "Fixed 141181" or "bug \#141181", but sometimes only the ID is
reported. Every positive integer number is a potential
issue.
To discern among issues and simple numbers we applied the following strategies:
\begin{enumerate}
    \item we considered only positive integer numbers present in the issue tracker as valid issue IDs related to the same release;
    \item we did not consider some numeric intervals particularly prone to be a false positive issue ID.
\end{enumerate}
The latter condition is not particularly restrictive in our study,
because we do not consider the first releases of the studied
projects, where issues with "low" ID appear.\\ All IDs not filtered
out are considered issues and associated to the addition or
modification of one ore more CUs, as reported in the commit logs. The
total number of issues hitting a CU in each release constitutes the
issue metric we consider in this study. Note that an issue reported
in an issue management system has a broad sense. It may denote an
error in the code, but also an enhancement of the system, or a
features request, or fixing a requirement error. Moreover, when many
CUs are affected by a single bug, it is possible that some of them
are modified not because they have the issue, but as a
side-effect of modifications made in other CUs.

\section{Results}\label{Results}
The subjects of our study were Eclipse and Netbeans projects,
both open source, object oriented, Java based systems. Table  \ref{tabella3.0.1} and \ref{tabella3.0.2} show the
number of CUs involved in the main releases of Eclipse and Netbeans, respectively.\\

\begin{table}
\caption{Number of CUs of Eclipse for each main release}
\begin{center}
\label{tabella3.0.1}
%\resizebox {0.48\textwidth }{!}{ %
\begin{tabular}{|l|c|c|c|c|c|}
\hline
Release         & 2.1   & 3.0  & 3.1  & 3.2  & 3.3     \\
\hline
Number of CU    & 7885  & 10584 & 12174 & 13221 & 14564\\
\hline
\end{tabular}%}
\end{center}
\end{table}

\begin{table}[htb!]
\begin{center}
\caption{Number of CUs of Netbeans for each main release}

\label{tabella3.0.2}
%\resizebox {0.49\textwidth }{!}{ %
\begin{tabular}{|l|c|c|c|c|c|c|c|}
\hline
Release         & 3.2    & 3.3   & 3.5   & 3.6   & 4.0  & 5.0   & 6.0    \\
\hline
Number of CU    & 3350   & 4421  & 7391  & 8350  & 9365 & 12137 & 37145    \\
\hline
\end{tabular}%}
\end{center}
\end{table}

A software system usually evolves through subsequent releases .
Main releases entail substantial enhancements of the
system, and are usually characterized by significant changes in
software sizes, as demonstrated by the data reported in Tables
\ref{tabella3.0.1} and \ref{tabella3.0.2}. Between two main releases
there may be different ``patching releases'', intended to fix bugs and to provide
minor enhancements. Even if we analyzed all the releases, we report
results for the main releases and the patching release immediately
preceding the next main release.
In fact most of bugs are introduced in upgrading from the last patching release
to the next main release.

\subsection{Statistical analysis}
\begin{figure*}[htbp]
%\begin{center}
    \begin{minipage}{0.4\textwidth}
    %\centering
    \includegraphics[scale=0.40]{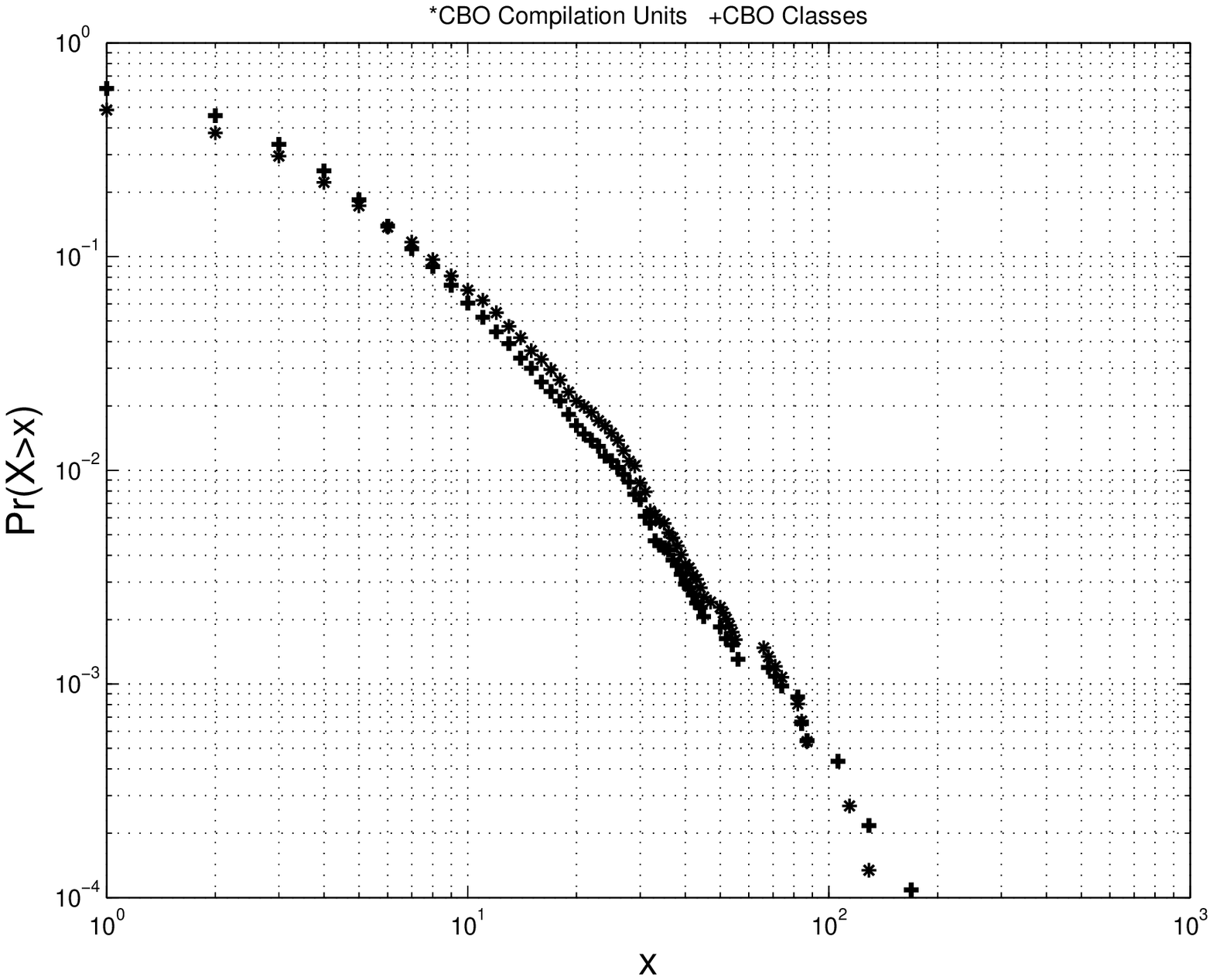}
    \caption{The CCDF of CBO metrics for classes (crosses) and CUs (stars) in Eclipse 3.2}\label{figure 3.1}
    \end{minipage}\hspace{20mm}
    \begin{minipage}{0.4\textwidth}
    %\centering
    \includegraphics[scale=0.40]{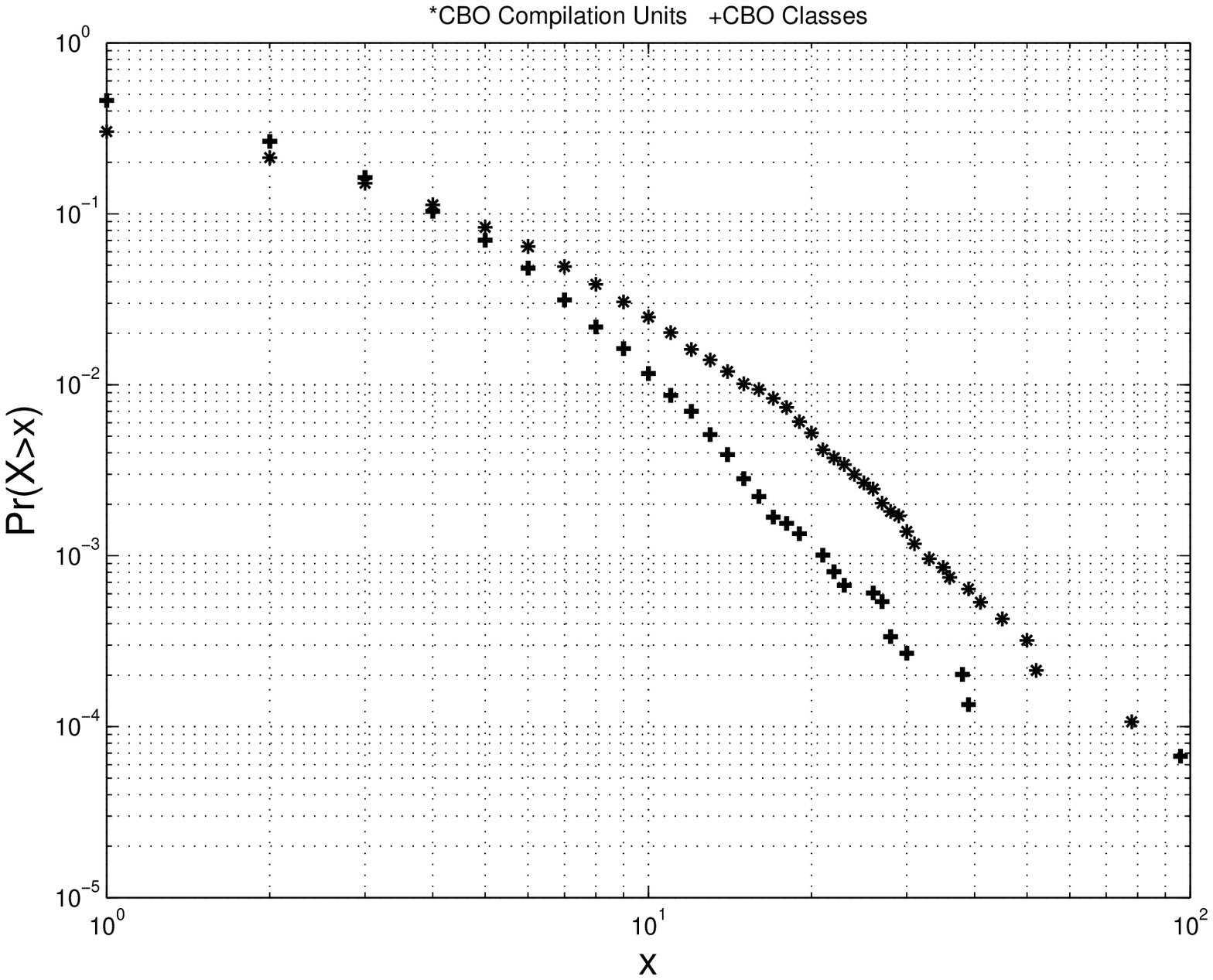}
    \caption{The CCDF of CBO metrics for classes (crosses) and CUs (stars) in Netbeans 4.0.}\label{figure 3.2}
    \end{minipage}
%\end{center}
\end{figure*}

We computed the statistical distributions of software metrics
underlying the software graph. We compared the metrics for software
graphs built using classes as basic units, already observed in
literature, with the ones obtained in this work for software graphs
built considering CUs.\\ The latter distributions substantially keep
the "fat-tail" behavior of the corresponding class metrics
\cite{Concas} in all cases. Fig. \ref{figure 3.1} reports the
log-log plot of the complementary cumulative distribution functions
(CCDF) of CBO metric of Eclipse 3.2 for classes and for CUs.

Fig. \ref{figure 3.2} reports the CCDF of CBO metrics, this time
referred to Netbeans 4.0. All these distributions exhibit a
power-law behavior in their tail. \\We recall that a quantity $x$
obeys a power law if it is drawn from a probability distribution
proportional to a negative power of $x$:
\begin{equation}\label{eq 3.1}
    p(x) \propto x^{-\gamma }~\text{where}~\gamma > 0. %falign
\end{equation}

$\gamma$ is the power-law coefficient, known also as the
\emph{exponent} or \emph{scaling parameter}. The corresponding
complementary cumulative distribution function (CCDF), i.e. the
probability that the random variable is greater than a given value
$x$, is:
\begin{equation}\label{eq 3.2}
    P(X\ge \ x) \propto x^{-(\gamma-1) }
\end{equation}

\begin{figure*}[htbp]
    \begin{minipage}{0.4\textwidth}\hspace{-10mm}
        \includegraphics[scale=0.45]{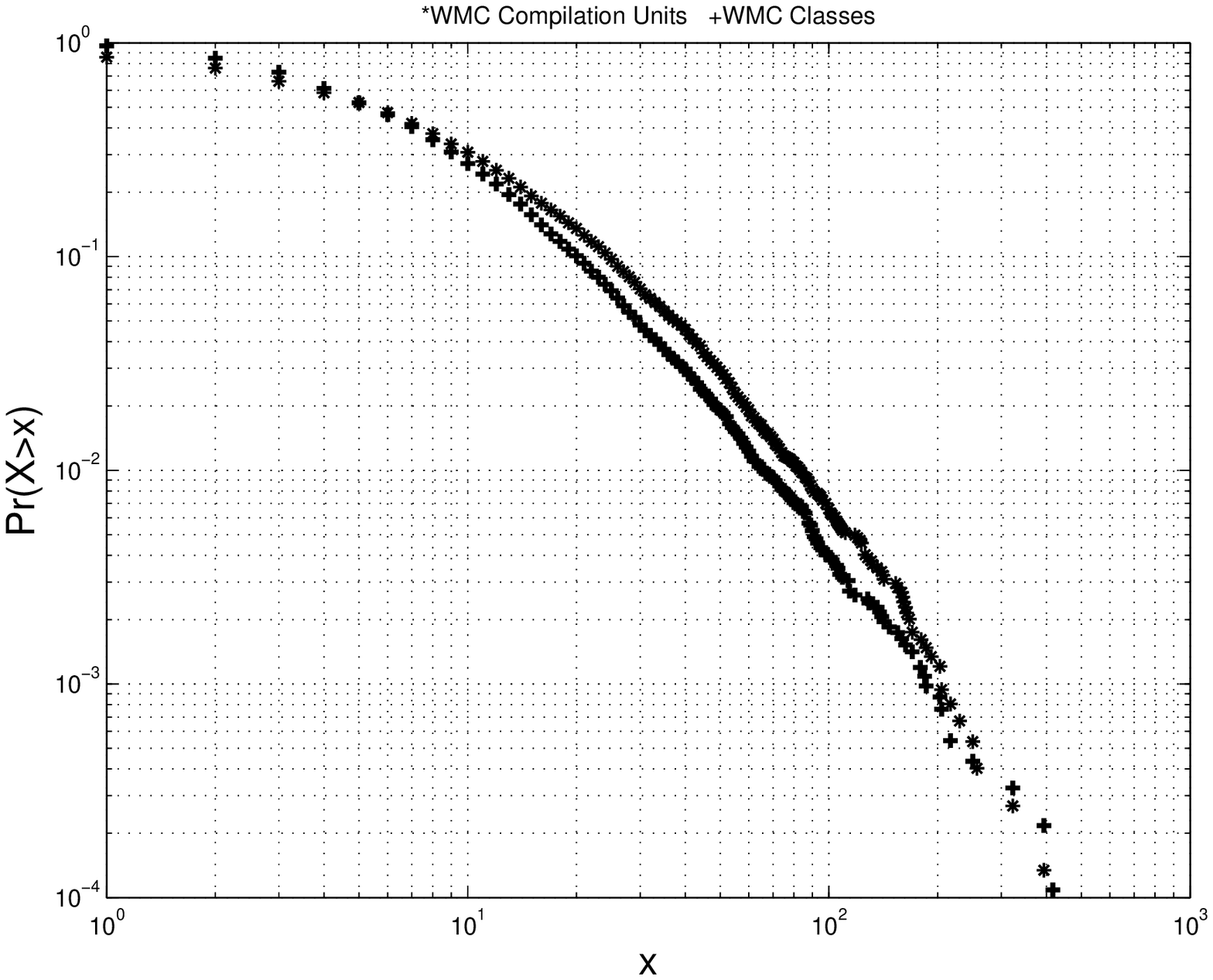}
        \caption{The CCDF of WMC metrics for classes (crosses) and CUs (stars) in Eclipse 3.2. }\label{figure 3.3}
    \end{minipage}
    \begin{minipage}{0.4\textwidth}
        \includegraphics[scale=0.45]{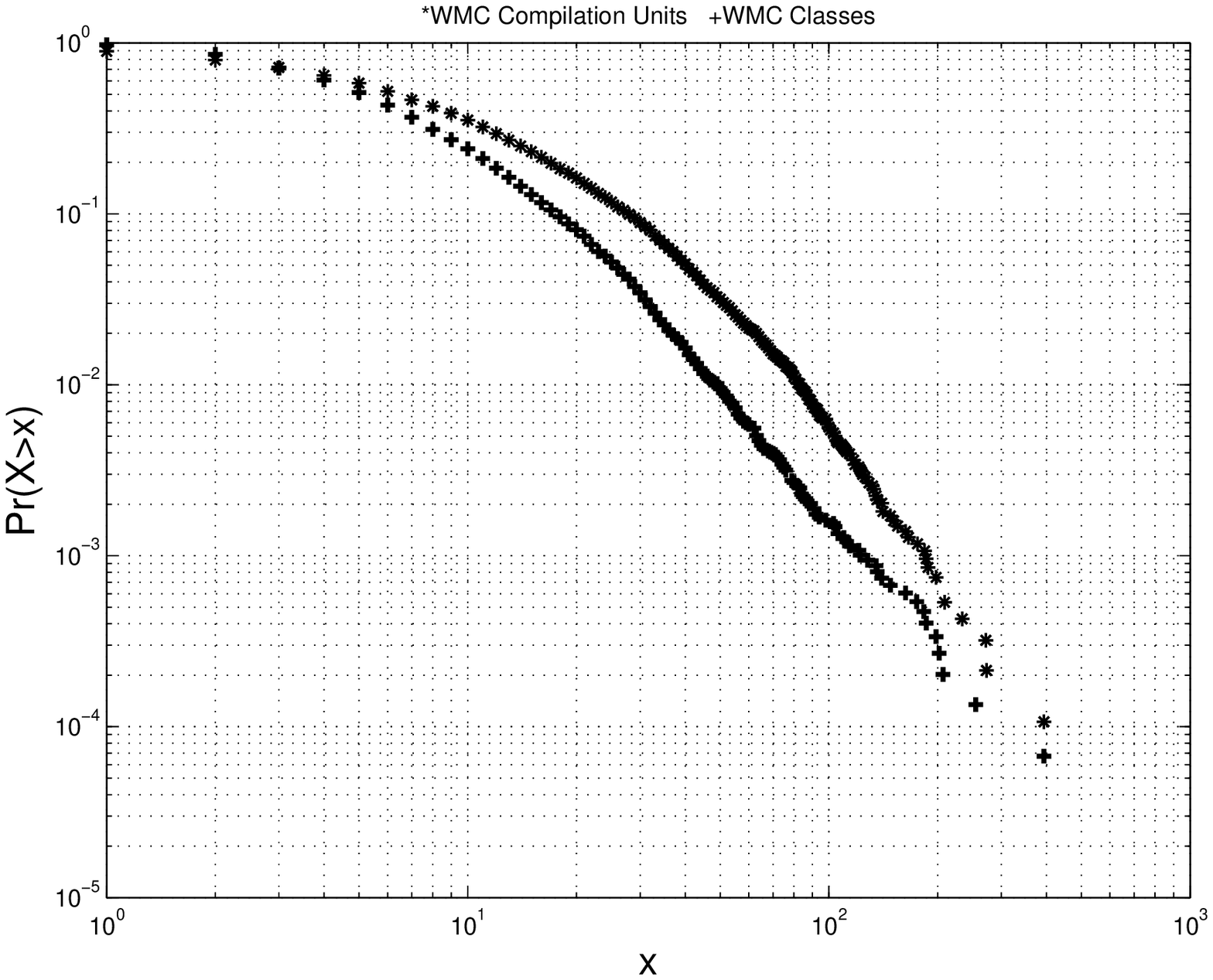}
        \caption{The CCDF of WMC metrics for classes (crosses) and CUs (stars) in Netbeans 4.0. }\label{figure 3.4}
    \end{minipage}
    \begin{minipage}{0.4\textwidth}\hspace{-10mm}
        \includegraphics[scale=0.45]{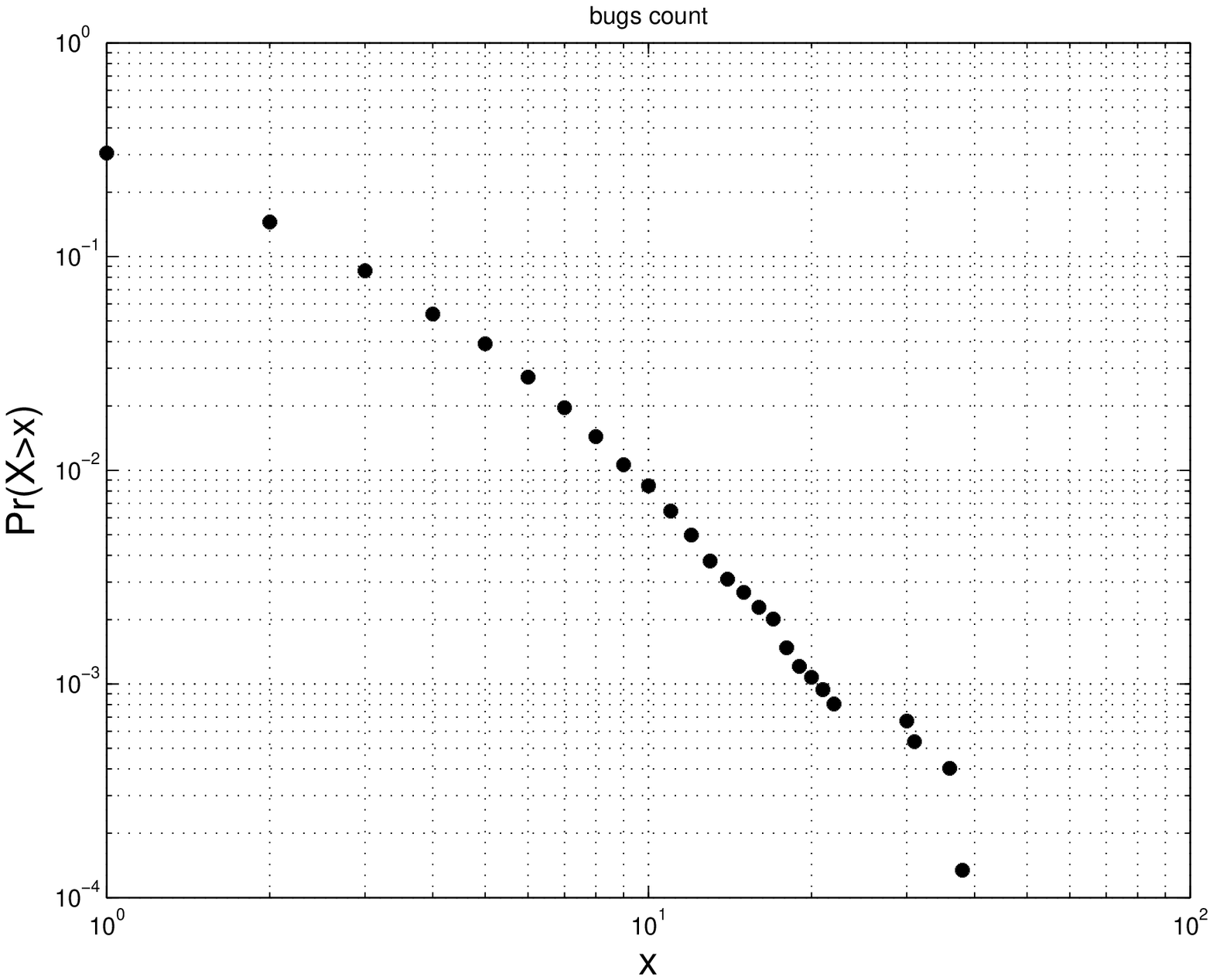}
        \caption{The CCDF of the number of bugs per CU in Eclipse 3.2. }\label{figure 3.5}
    \end{minipage}\hspace{20mm}
    \begin{minipage}{0.4\textwidth}
        \includegraphics[scale=0.45]{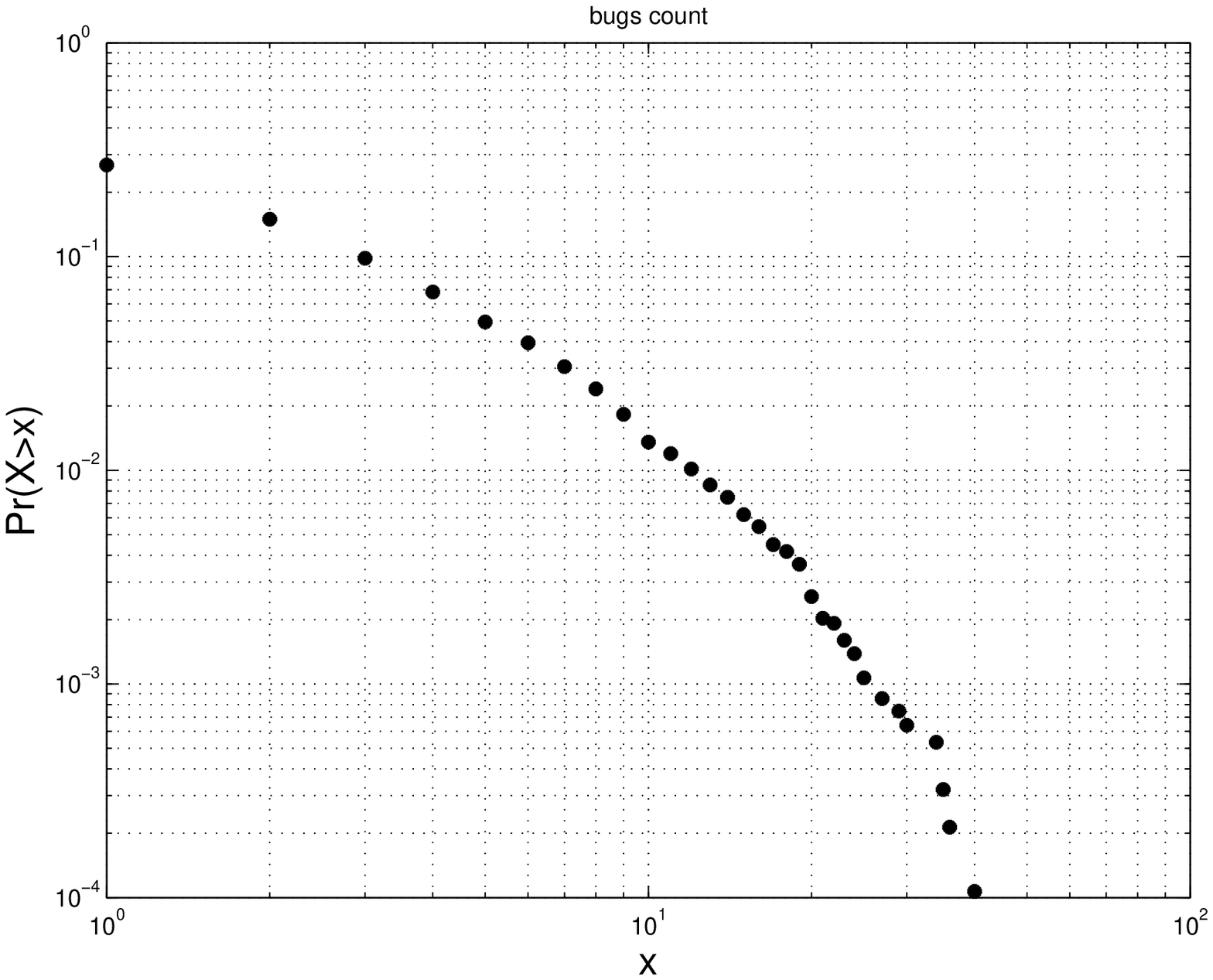}
        \caption{The CCDF of the number of bugs per CU in Netbeans 3.4. }\label{figure 3.6}
    \end{minipage}%\hfill
\end{figure*}

A power-law, or Pareto, distribution cannot hold for $x$ = 0, so
eligible values of $x$ must be greater than a positive number
$x_{min}$. This characteristic allows to consider distributions that
are power-laws only in their right "tail", that is for $x$ greater
than a given value $x_{min}$, and not for lower
values of $x$. All the distributions shown in Figs. \ref{figure 3.1},
\ref{figure 3.3} and \ref{figure 3.4} show a straight line
behavior in their right tail.
Note that the CCDF has the same analytical expression of
the distribution function, with a negative exponent offset by one.
Plotting $p(x)$ or $P(x)$ in log-log scale one obtains a straight line, as shown in Figs.
\ref{figure 3.1} and \ref{figure 3.2}.  \\ Fig. \ref{figure 3.3} and
\ref{figure 3.4} show the CCDF of WMC metric in Eclipse 3.2 and in
Netbeans 4.0, respectively. These distributions are also quite
similar, and present again in their tail a power-law behavior, both
for classes and for CUs. We found this behavior also for all other
releases, and for all metrics. \\ The finding that the distributions
of CU metrics largely coincide with those of the corresponding
metrics of classes suggests that the same considerations that are
valid for CUs may be extended also to classes, even in the cases
where data for the classes are not directly accessible, like in our
case for bugs. One goal of this paper is, in fact, to find, by means
of the software graph framework, existing correlations among bugs
and metrics. Thus, since bug information for classes is not
directly detectable from the repository, we analyzed the
bugs metric only for CUs, and use this information to obtain clues
about classes. \\ Fig. \ref{figure 3.5} shows the CCDF of the number
of bugs per CU in Eclipse 3.2. Fig. \ref{figure 3.6} shows the same
distribution in Netbeans 3.4. The meaning of these power-law tail
distributions is unequivocal. While most CUs present only very few
bugs, there is a non-negligible number of CUs with very many bugs.
We also found similar shapes (patterns) in all other main
releases.

On the basis of these similarities, the hypothesis that the
power-laws existing for bug distribution among CUs may be extended
to classes, as well as to other units, like modules or packages, and
that it is a property of the graph structure of the system looks
sensible.

In fact similar
results were obtained by Andersson et Runeson \cite{Andersson}, and
by Zhang \cite{Zhang}.
Andersson et Runeson suggest a Pareto law governing the distribution
of bugs across basic units of a software system only partially OO, showing that few
modules contain most of the bugs (the 20-80 rule \cite{Juran}).
Zhang re-examined their results for the Eclipse software system,
finding that a Weibull distribution fits data better than a
power-law, studying packages instead of modules. Since
the tail of a Weibull distribution is often not distinguishable
from a power-law tail, their results support our hypothesis.\\

\begin{figure*}[htbp]
    \begin{minipage}[c]{0.4\textwidth}
        \includegraphics[scale=0.4]{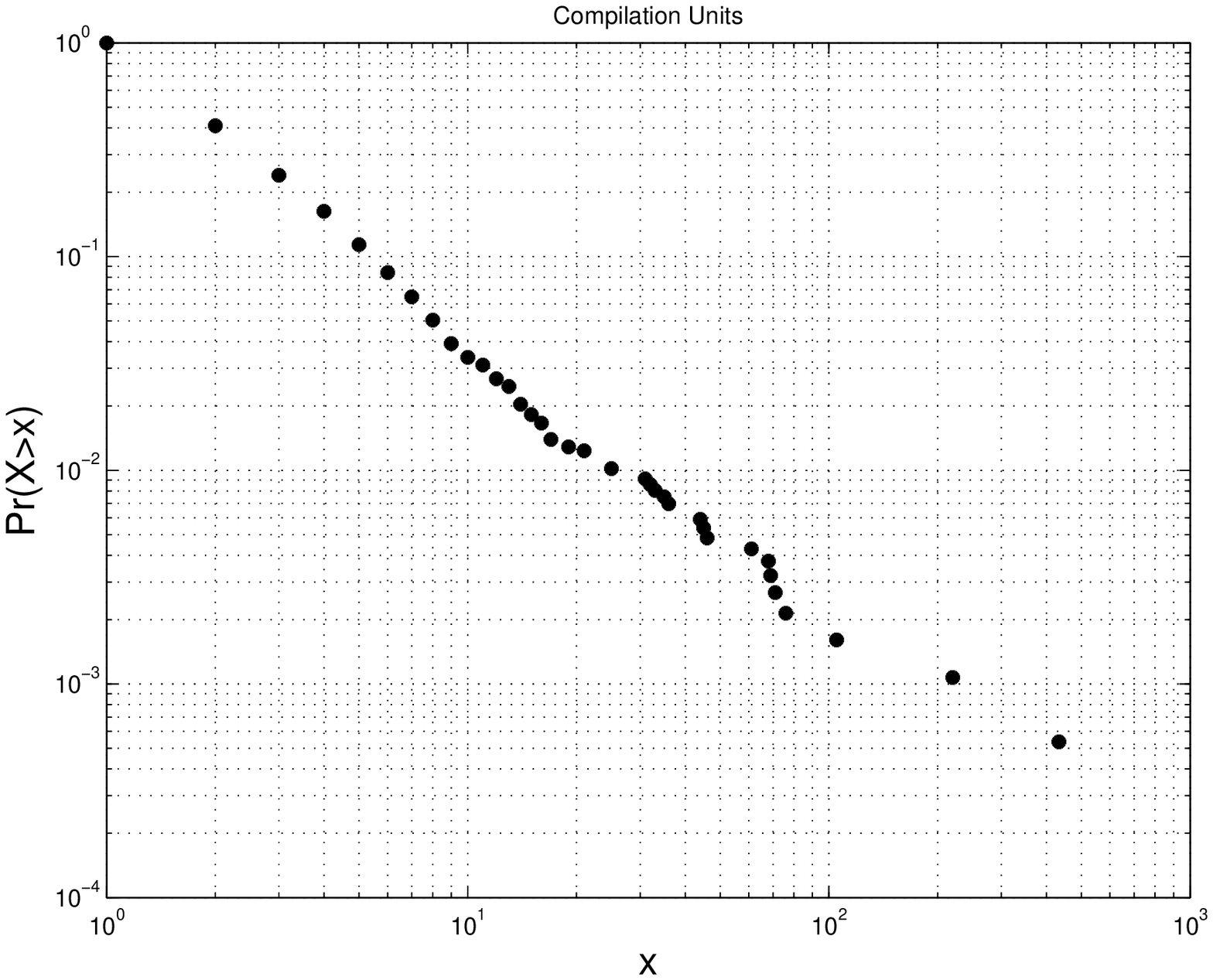}
        \caption{The CCDF of the number of CUs associated to each bug in Eclipse 3.2.}\label{figure 3.7}
    \end{minipage}\hspace{20mm}
    \begin{minipage}[c]{0.4\textwidth}

        \includegraphics[scale=0.4]{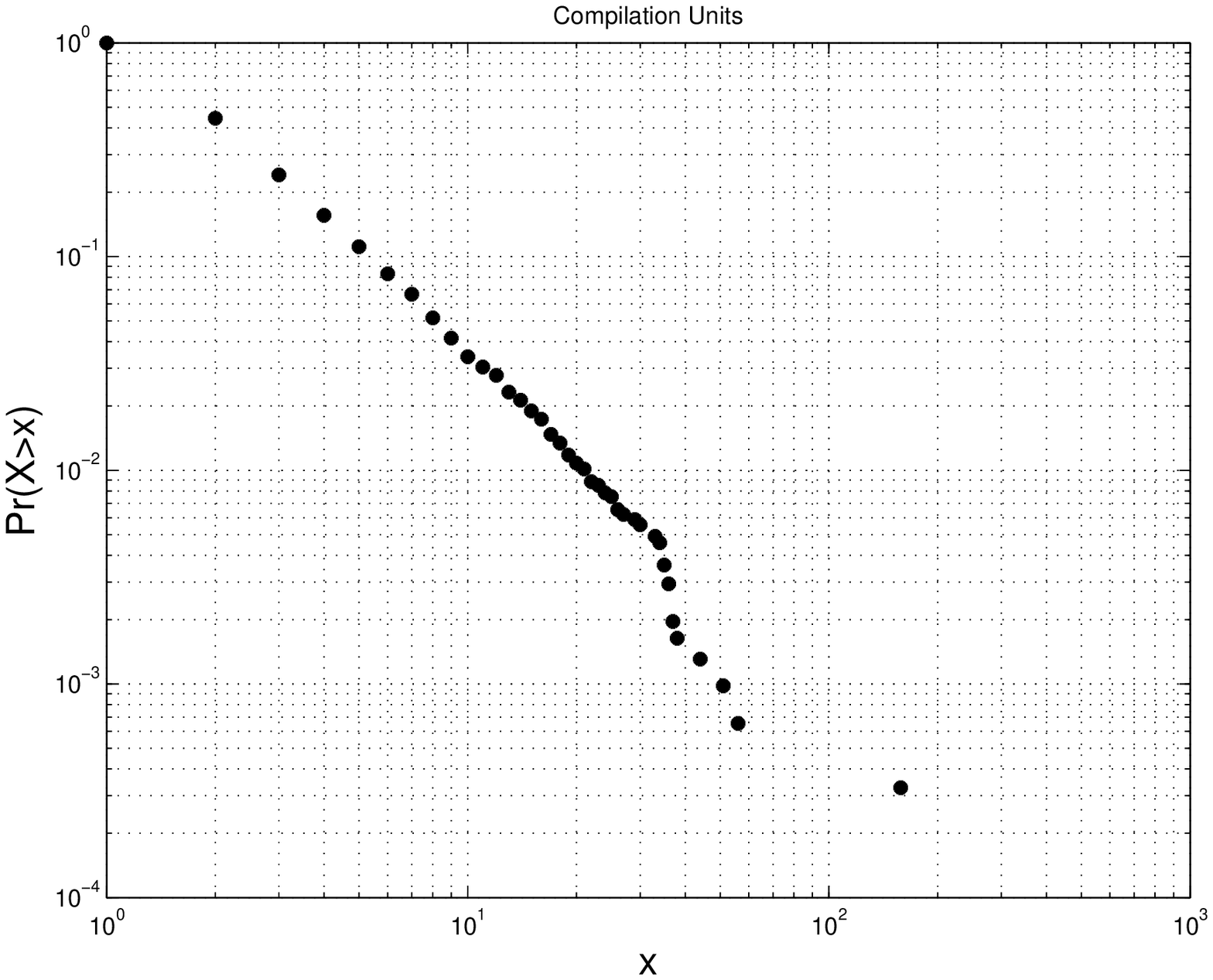}
        \caption{The CCDF of the number of CUs associated to each bug in Netbeans 3.4.}\label{figure 3.8}
    \end{minipage}
\end{figure*}

Let us point out what we consider our most relevant finding. We
verified that a power-law distribution may be appropriate to
describe the fat-tail distribution of different quantities. Note
that the fat-tail contains the software units
to which
most of the information belongs. When a metric is distributed according to
a power-law, even only in its tail, with a scaling exponent small enough,
there are relatively few units with highest values of the metrics, where
criticality resides, while most other units are much less critical. The 80-20
Pareto principle is a consequence of that: about 80\% of the criticality is held
in 20\% of all units.\\
Our analysis is finer than those performed in \cite{Andersson} or in
\cite{Zhang}, in the sense that we analyzed the software structure
and relationships at the level of compilation units, one level
deeper than the module or the package level presented in the above
works. This allowed us to recover finer information on the
distributions of metrics, especially in their tail. Our results
confirm those of Andersson and Runeson, and of Zhang, showing that
the same framework holds at different scales, exhibiting a
scale-free structure \cite{Barabasi}. This finding qualitatively
supports the use of power-laws. Finally, also Louridas et al.
\cite{Louridas}, show a large variety of cases in which power-laws
well account for the distribution of different software properties.
\\ Regarding the value of the exponent $\gamma$ and the
corresponding behavior of the number of bugs per CU, this value
tends to be between 2.5 and 3.5 in the various releases examined for
both Eclipse and Netbeans.
\\According to ref. \cite{Louridas}, a mathematical
description of the fat-tail may have relevant consequences on
software engineering, for example in helping to carefully select
which parts of the software project are worth of
more care and effort, also from an economical point of view. \\
For instance, given $n$ modules characterized by a metric
distributed according to a power-law with exponent $\gamma$,
the average maximum expected value for this metric in
the module with highest metric value, $<xmax>$, is given by the
formula \cite{Newman}
\begin{equation}\label{eq Newman}
    <x_{max}> = n^{1/(\gamma-1)}
\end{equation}
This formula provides a definite expectation of the maximum value
taken by the metric, and hence allows to flag specific modules with
metric value of this order of magnitude.\\
We studied also the distributions of the number of CUs hit by a
single bug, the dual of the distribution of bugs across CUs. Also in
this case, we find a power-law, as shown in Figs. \ref{figure 3.7}
and \ref{figure 3.8} for Eclipse and Netbeans, respectively. This
means that, while most bugs affect just one or a few CUs, there are
bugs that affect tens, or ever hundreds of CUs.

The value of the exponent $\gamma$ of the distributions of the
number of CUs affected by a bug is consistently between 2.2 and 2.9
in all considered releases, for both Eclipse and Netbeans, meaning
an ever ``fatter'' tail of this distribution with respect to the
previously studied distribution of bugs per CU. \\
The finding that the distribution of bugs across CUs satisfies a
power-law, may suggest a model for the introduction and the spread
of bugs in the software system. We already specified that, in our
investigation, we name ``bug'' each numerical identifier found in
the repository associated to software ``fixing''. Thus, generally
speaking, a bug reported in a CU means that such a CU needed to be
partially modified owing to this bug. Now, let us consider the graph
structure of the software system. We, and many other authors in
literature, verified an organized structure of such graphs,
exibiting power-law distributions for many properties of the system.
In particular, there are nodes
linked with many other nodes, playing the role of  "hubs" of the
system. For example, there are few CUs with a large number of
in-links, meaning that they are extensively used by other CUs. If a
bug hits such CUs, namely, the CU code need modifications, it is
very likely that also the code of CUs linked to that node need to be
modified. Such mechanism may generate a sort of defect propagation
in the software graph, very similar to the spread of a contagious
disease. The system gets infected by bugs, and a single bug may
affect many different CUs, if it propagates from a hub node. On the
contrary, bugs in CUs with very few links will
likely remain confined to a small number of CUs. \\
Our heuristic conclusion is that the power-laws observed for the bug
distribution is probably due to the scale-free structure of the
software graph. Bugs propagate inside a constraining framework,
which determines their diffusion across the software system. \\
From the software engineering point of view, the usefulness of
finding power-laws in the tail of the bugs distribution, may be
illustrated following the reasoning of Louridas et al.
\cite{Louridas}. Once it is shown that bugs distribution across CUs
is in the form of a power-law, CUs in the tail may be identified as
the most fault-prone. Thus, after the issue of a new release, the
inspection of CUs for bug detection may take advantage of this
information. For instance, an inspection of the highest 5 \% ranked
CUs would imply the inspection of a high percentage of bugs, were
the exact percentages is related to the power-law exponent.

\subsection{Correlation}
We analyzed, for each version of the system, the correlations
between the considered software metrics and the number of bugs. This
information may be used to understand, from the measure of the
metric, which parts of the software are most affected by faults, and
to devise the possible strategies to apply during software
development in order to control metrics values, with the goal of
reducing bug introduction. \\
Our analysis started computing, for various releases $R_{i}$ of the
system, the linear correlation between a particular CK metric and
the number of bugs of the same CUs. This is only a preliminary
analysis in order to identify which CU metrics are more related to
fault proneness.
We recall that developers distinguish between "main" and "patching"
releases, and that changes from a main release to the next are
usually relevant also regarding metrics. \\
In the first part of our study we referred to the main releases. In
the Eclipse project main releases are identified by two-digit
numbers, that is: Eclipse 2.1, Eclipse 3.0, Eclipse 3.1, Eclipse
3.2, and Eclipse 3.3. We analyzed what can be deduced about bugs
from the analysis of the software metrics for this kind of releases.\\
Table \ref{tabella3.2.1} shows the correlations between metrics and
bugs for the main releases of Eclipse. The metrics showing the
highest correlation with bugs are those taking into account the
number of dependencies with other CUs, namely CBO and RFC. This fact
highlights the importance of an analysis of a software system as a
graph. The out-links metric is less correlated with bugs than CBO
and RFC. Out-links metric includes not only dependency
relationships, but also inheritance and implements relationships. A
lower correlation of this metric with bugs may be interpreted with a
higher ability of dependency relationships of propagating bugs with
respect to the other relationships.

\begin{table}[h]
\begin{center}
\caption{Pearson correlations between metrics and bugs for some
releases of Eclipse:} \label{tabella3.2.1}
\begin{tabular}{|l|r|r|r|r|r|}
\hline
            & 2.1   & 3.0  & 3.1  & 3.2  & 3.3     \\
\hline
bugs-LOCS       & 0.49  & 0.57 & 0.54 & 0.58 & 0.48\\
\hline
bugs-CBO        & 0.55  & 0.53 & 0.55 & 0.55 & 0.42\\
\hline
bugs-RFC    & 0.59  & 0.48 & 0.44 & 0.56 & 0.45\\
\hline
bugs-WMC    & 0.48  & 0.45 & 0.38 & 0.48 & 0.40\\
\hline
bugs-LCOM   & 0.30  & 0.21 & 0.15 & 0.34 & 0.24\\
\hline
bugs-inliks & 0.1   & 0.17 & 0.25 & 0.28 & 0.24\\
\hline
bugs-outlinks   & 0.47  & 0.38 & 0.40 & 0.55 & 0.42\\
\hline
\end{tabular}
\end{center}
\end{table}

The low correlation of the in-links metric with bugs indicates that
it is important to take into account not only the number of links
but also their direction. An out-link directed from a compilation
unit A to a compilation unit B may be considered like a channel
easing the propagation of defects from B to A, but not vice-versa.\\
Another metric that is well correlated with bugs is LOCS metric.
This result can be clearly understood considering that LOCS metric
is well correlated with CBO, RFC and out-links metrics. Moreover,
the larger the CU, the higher the probability of being hit by some
bugs. The LCOM metric is calculated taking into account the internal
structure of a compilation unit, and not the relationships with
other CUs. The low correlation between LCOM and bugs suggests that
the fault proneness of a CU is not overly influenced by the lack of
cohesion of the classes contained in the CU. This confirms once
again the relevance of the information provided by an analysis of
the software system viewed as a graph. The data show that, while
there are metrics more or less correlated with the number of bugs,
the correlations are never very strong. This is sensible, since a
perfectly linear correlation would imply, for example, a doubling of
the introduced bugs with the doubling of the metric, and this never
occurs in reality.

\begin{table}[h]
\begin{center}
\setlength{\parindent}{0pt} \caption{Pearson  correlations between
metrics and bugs for all releases of Netbeans} \label{tabella3.2.2}
\begin{tabular}{|l|r|r|r|r|r|r|r|}
\hline
               & 3.2   & 3.3  & 3.5 & 3.6  & 4.0  & 5.0 & 6.0     \\
\hline
bugs-LOCS      & 0.34  &0.55  &0.42 & 0.4  & 0.36 &0.34 & 0.35    \\
\hline
bugs-CBO       & 0.25  &0.44  &0.37 & 0.36 & 0.27 &0.28 & 0.25    \\
\hline
bugs-RFC       & 0.38  &0.57  &0.44 & 0.39 & 0.33 &0.31 & 0.28    \\
\hline
bugs-WMC       & 0.38  &0.53  &0.38 & 0.35 & 0.31 &0.27 & 0.23    \\
\hline
bugs-LCOM      & 0.32  &0.44  &0.23 & 0.13 & 0.10 &0.08 & 0.04    \\
\hline
bugs-inliks    & 0.10  &0.16  &0.19 & 0.12 & 0.05 &0.07 & 0.07    \\
\hline
bugs-outlinks  & 0.24  &0.40  &0.35 & 0.33 & 0.24 &0.25 & 0.23    \\
\hline
\end{tabular}
\end{center}
\end{table}

In Table \ref{tabella3.2.2} we report the correlation between a
metric and the number of bugs of the CUs for various releases
$R_{i}$ of the Netbeans system. In Netbeans the distinction between
main and patching releases is fuzzier than in Eclipse; moreover there are various MR which are not followed by classic PR.\\
A comparison of Tables \ref{tabella3.2.1} and \ref{tabella3.2.2}
shows that Netbeans correlation values among metrics and bugs number
are usually lower than in Eclipse. However, in both systems, LOCS
and RFC are the two most correlated metrics to the CU faultness,
while LCOM shows, in both cases, a weak correlation to CU
faultness.\\
These results show that:
\begin{itemize}
  \item Given a release, there exist metrics that are more correlated to CU faultness than
  others;
  \item Considering all releases, there is not one CK metric which is the most correlated for each release;
  \item Given a metric, its correlation with the number of bug changes release by release.
\end{itemize}

Note, however, that all correlation coefficients shown in Table
\ref{tabella3.2.1} and \ref{tabella3.2.2} are positive, so all the
considered metrics are, more or less, positively correlated with
bugs. This is consistent with the observation that all CK metrics
and the size of the code are a measure of complexity, and
therefore should in general be kept low.

\subsection{Analysis of software
evolution} We also analyzed the evolution of the metrics between two
consecutive releases.
To this purpose we define different types of CUs, distinguishing among
updated, unmodified, newly introduced, and defining all these types
with respect to all the different metrics.\\
In particular, given a release $R_{i}$, the next release $R_{i+1}$,
and a metric M, we classified the compilation units in four
categories:
\begin{itemize}
  \item CU.X is the set of compilation units where metric M doesn't change between $R_{i}$ and $R_{i+1}$;
  \item CU.U is the set of compilation units where metric M changes (Updated);
  \item CU.A is the set of compilation units that exist in $R_{i+1}$ but not in $R_{i}$ (Added);
\end{itemize}
It must be pointed out that U and X categories are defined relative
to a specific metric. A CU might exhibit a change in metric M but
not in metric M' between the releases $R_{i}$ and $R_{i+1}$. Thus,
it will belong to class CU.U for M, and to class CU.X for M'. This
case is not common, but it is definitely possible. CU.A is defined
regardless to any metric M, since it refers to CUs just introduced
in the new release. There are also CUs existing in release $R_{i}$
but not in release $R_{i+1}$. These deleted CUs are not considered
in our study.\\
Given the set of compilation units belonging to the three categories
CU.U, CU.X, and CU.A, we compute:
\begin{itemize}
  \item the fraction of  compilation unit affected by bugs, which provides an infection probability;
  \item the average number of  bugs of the infected compilation units.
\end{itemize}
In Table \ref{tabella3.2.3} we show the probability for CUs
belonging to one of the families U, X and A, of being infected, in
various changes of releases.
\begin{table}[h]
\begin{center}
\caption{Percentage of bug-affected CUs between two consecutive
releases (shown in the top row), for different families, relative to
different metrics in Eclipse} \label{tabella3.2.3}
\begin{tabular}{|l|r|r|r|r|r|}  \hline

\multicolumn {1}{|c}{}& \multicolumn {1}{c}{} &
\multicolumn{3}{|c|}{Subsequent releases} \\  \hline Metric
&Set & 2.1.3-3.0 &3.0.2-3.1  &  3.2.2-3.3 \\  \hline
\multirow{2}{*}{LOC}  &CU.U & 0.66 &0.61  &  0.62 \\ \cline{2-5}
                      &CU.X & 0.15 &0.17  &  0.1  \\ \hline
\multirow{2}{*}{CBO}  &CU.U & 0.6  &0.7   &  0.68 \\ \cline{2-5}
                      &CU.X & 0.2  &0.27  &  0.18 \\ \hline
\multirow{2}{*}{LCOM} &CU.U & 0.7  &0.69  &  0.66 \\ \cline{2-5}
                      &CU.X & 0.22 &0.23  &  0.16 \\ \hline
                      \hline
\multicolumn {2}{|c|}{CU.A} & 0.51 &0.55  &  0.58 \\ \hline
\end{tabular}
\end{center}
\end{table}

The probability that a CU belonging to family CU.U is infected is
between 0.6 - 0.7 in Eclipse. This means that there is a high
probability that changing the LOCS, CBO, or LCOM metrics of a CU
from one release to the next results in injecting at least one error
into the compilation unit. This result confirms Purushothaman's
study \cite{Purushothaman}, that highlighted that code correction
for defects often introduces new defects. Also the CUs added to the
system, in the transition from $R_{i}$ to $R_{i+1}$, show a high
probability to be infected, clearly larger than for the case
of CUs not modified (set CU.X), and slightly smaller than for the set
CU.U. Similar results were obtained also for all other
metrics.\\
On the contrary, if the metric does not change there is a low
probability that a CU is affected by bugs. These bugs clearly refer
to bugs already present in $R_{i}$ but that were found only when
checking $R_{i+1}$ release. \\
In order to support our findings about the deep differences among
CU.U, CU.X and CU.A families, we performed chi-square significance
tests. We formulate the following null hypothesis: ``the subdivision
of CU in U, X and A does not significantly influence
the number of infected CU''.\\
We verified that all $\chi^2$ values have a confidence level larger
than 99.9 percent
 (the confidence level is actually much larger). Therefore we can reject the null hypothesis with a probability greater
  than 99.9\%, and confirm that our classification of CUs into families provides significative correlations with the
   presence of bugs.

In Table \ref{tabella3.2.5} we report the average number of bugs of
the infected CUs. These data confirm that
 the CUs infected of type U and A have an average number of bugs larger than the compilation units of type X.
  Note also that, on average, more than one bug is found during a release lifespan even in the CUs that are
  not changed in the release.
\begin{table}[h]
\begin{center}
\caption{Average number of bug-affected CUs between two consecutive
releases (shown in the top row), for different families, relative to
different metrics in Eclipse} \label{tabella3.2.5}
\begin{tabular}{|l|r|r|r|r|r|}  \hline

\multicolumn {1}{|c}{}& \multicolumn {1}{c}{} &
\multicolumn{3}{|c|}{Subsequent releases} \\  \hline Metric
&Set & 2.1.3-3.0 &3.0.2-3.1  &  3.2.2-3.3 \\  \hline
\multirow{2}{*}{LOC}  &CU.U & 4.02 &3.16  &  2.61  \\ \cline{2-5}
                      &CU.X & 1.38 &1.22  &  1.29  \\ \hline
\multirow{2}{*}{CBO}  &CU.U & 3.92 &3.88  &  3.03  \\ \cline{2-5}
                      &CU.X & 2.36 &1.86  &  1.8   \\ \hline
\multirow{2}{*}{LCOM} &CU.U & 4.34 &3.58  &  2.95  \\ \cline{2-5}
                      &CU.X & 2    &2.64  &  1.66  \\ \hline
                      \hline
\multicolumn {2}{|c|}{CU.A} & 3.2 &2.73  &  2.51 \\ \hline
\end{tabular}
\end{center}
\end{table}
Thus, in general,irrespectively of the metric,~we~have:
\begin{itemize}
  \item CU.U infection probability is around 60-70\%;
  \item CU.A infection probability is around 50-60\%;
  \item CU.X infection probability is around 10-30\%;\\
\end{itemize}

CU.U are the most faultprone, followed by CU.A. The mean number of
bug is in agreement with these results, and varies between:
\begin{itemize}
  \item 2.5 and 4 for CU.U;
  \item 2.5 and 3.5 for CU.A;
  \item 1.6 and 2.5 for CU.X.
\end{itemize}

In Table \ref{tabella3.2.6} we show the results for Netbeans. In
Netbeans there are less PRs, thus we consider
jump of releases between couples of MRs. As in
Eclipse, also in Netbeans LOC, CBO or LCOM variations determine a
major introduction of bugs into the system, whereas the addition of new
CUs determines a slightly lower rate of bug injection. Similar
results were obtained also for the other metrics.

\begin{table}[h]
\begin{center}
\caption{Percentage of infected CUs between two consecutive releases
(shown in the top row), for different families relative to different
metrics in Netbeans.} \label{tabella3.2.6}
\begin{tabular}{|l|r|r|r|r|r|}  \hline

\multicolumn {1}{|c}{}& \multicolumn {1}{c}{} &
\multicolumn{4}{|c|}{Subsequent releases} \\  \hline Metric
&Set & 3.1-3.2 &3.2.1-3.3  &  3.6-4.0 & 4.1-5.0 \\  \hline
\multirow{2}{*}{LOC}  &CU.U & 0.68 &0.67  &  0.62 &0.61 \\
\cline{2-6}
                      &CU.X & 0.19 &0.15  &  0.07 &0.06 \\ \hline
\multirow{2}{*}{CBO}  &CU.U & 0.51 &0.57  &  0.59 &0.67 \\
\cline{2-6}
                      &CU.X & 0.27 &0.31  &  0.18 &0.15 \\ \hline
\multirow{2}{*}{LCOM} &CU.U & 0.53 &0.58  &  0.67 &0.6 \\
\cline{2-6}
                      &CU.X & 0.23 &0.21  &  0.12 &0.11 \\ \hline
                      \hline
\multicolumn {2}{|c|}{CU.A} & 0.47 &0.28  &  0.36 &0.39 \\ \hline
\end{tabular}
\end{center}
\end{table}

The chi-square significativity test, about the classification in
families for the Netbeans
projects, performed using the same null hypothesis used for Eclipse
yielded again confidence
levels higher than 99.9 percent.\\
Table \ref{tabella3.2.7} shows bug mean values for different CUs
families. Again, updated and added CUs show higher mean values than
unchanged CUs. In this case,
 CU.U and CU.A show values closer than in Eclipse. Also in Netbeans, on average, more than
  one bug is found during a release lifespan even in the CUs that are not changed.

\begin{table}[h]
\begin{center}
\caption{Average number of bugs of the infected CUs relative to
different metrics in Netbeans.} \label{tabella3.2.7}
\begin{tabular}{|l|r|r|r|r|r|}  \hline

\multicolumn {1}{|c}{}& \multicolumn {1}{c}{} &
\multicolumn{4}{|c|}{Subsequent releases} \\  \hline Metric
&Set & 3.1-3.2 &3.2.1-3.3  &  3.6-4.0 & 4.1-5.0 \\  \hline
\multirow{2}{*}{LOC}  &CU.U & 3.52 &3.66  &  2.69 &2.62 \\
\cline{2-6}
                      &CU.X & 1.51 &1.44  &  1.29 &1.24 \\ \hline
\multirow{2}{*}{CBO}  &CU.U & 3.36 &3.75  &  3.65 &3.87 \\
\cline{2-6}
                      &CU.X & 2.14 &2.15  &  1.89 &1.85 \\ \hline
\multirow{2}{*}{LCOM} &CU.U & 3.28 &3.53  &  3.01 &3.14 \\
\cline{2-6}
                      &CU.X & 1.92 &1.64  &  1.58 &1.53 \\ \hline
                      \hline
\multicolumn {2}{|c|}{CU.A} & 2.62 &3.08  &  3.92 &3.35 \\ \hline
\end{tabular}
\end{center}
\end{table}

Summarizing these results, we found that:
\begin{itemize}
  \item the most infected CUs, in both projects, are updated CUs;
  infection probabilities values are almost 70\% in both systems;
  \item CUs belonging to CU.A set exhibits in general a slightly smaller infection probability than CU.U set;
  \item CUs belonging to CU.X set are much less infected than CUs belonging to CU.A, and never exceed 30\%
  probability to be hit by a bug;
  \item usually, updated CUs have more bugs than others; this is always true in Eclipse, whereas it is almost always true
  in Netbeans;
  \item In Eclipse, the mean number of bugs of CU.U sets is often higher than in Netbeans, whereas the opposite holds for CU.A set.
\end{itemize}

One of the main differences between Eclipse and Netbeans projects is
the clear subdivision between patching release
 and main release. In Eclipse it is simple to verify that each main release X.0 is always followed by patching releases,
  of type X.0.1, X.0.2, and so on. This distinction is weaker in Netbeans, and this seems to affect the variation
  of its statistics.\\
For the family of compilation units U (CU.U), we calculated the
correlation between the fractional change of
 some metrics, passing from $R_{i}$ to $R_{i+1}$ releases, and the number of bugs in $R_{i+1}$. We were
  interested in determining if and how the growth of a metric is possibly associated to an increase in the number of bugs.\\
In Tables \ref{tabella3.2.8} and \ref{tabella3.2.9} we report this
correlation for Eclipse and Netbeans projects.

\begin{table}[h]
\begin{center}
\caption{Pearson correlation between metric changes and number of
defect in the subsequent release in Eclipse.} \label{tabella3.2.8}
\begin{tabular}{|l|r|r|r|}  \hline

\multicolumn {1}{|c}{}&  \multicolumn{3}{|c|}{Subsequent releases}
\\  \hline Metric variation      & 2.1.3-3.0 &3.0.2-3.1  &
3.2.2-3.3  \\  \hline $\Delta$CBO-bugs      & 0.37 &0.58  &  0.49
\\ \hline $\Delta$LOCS-bugs     & 0.29 &0.64  &  0.53  \\ \hline
$\Delta$RFC-bugs      & 0.39 &0.56  &  0.51  \\ \hline
$\Delta$LCOM-bugs     & 0.33 &0.32  &  0.49  \\ \hline
\end{tabular}
\end{center}
\end{table}

\begin{table}[h]
\begin{center}
\caption{Pearson correlation between metric changes and number of
defect in the subsequent release in Netbeans} \label{tabella3.2.9}
\begin{tabular}{|l|r|r|r|r|}  \hline

\multicolumn {1}{|c}{}&  \multicolumn{4}{|c|}{Subsequent releases}
\\  \hline Metric variation      & 3.1-3.2 &3.2.1-3.3  &  3.6-4.0 &
4.1-5.0 \\  \hline $\Delta$CBO-bugs      & 0.18 &0.39  &  0.19 &0.55
\\ \hline $\Delta$LOCS-bugs     & 0.30 &0.61  &  0.55 &0.59 \\
\hline $\Delta$RFC-bugs      & 0.37 &0.60  &  0.56 &0.59 \\ \hline
$\Delta$LCOM-bugs     & 0.28 &0.68  &  0.45 &0.33 \\ \hline
\end{tabular}
\end{center}
\end{table}

All data in Tables \ref{tabella3.2.8} and \ref{tabella3.2.9}, show
positive correlations. Correlation values are quite similar for the
same pair of
 subsequent releases, whereas they show larger fluctuations for different metrics. A comparison of Tables \ref{tabella3.2.8} and \ref{tabella3.2.9}
  shows that correlation values in Netbeans are often lower than in Eclipse. This result can be partially due to the
   less clear subdivision between main and patching releases in Netbeans project.\\
According to tables \ref{tabella3.2.3}, \ref{tabella3.2.6},
\ref{tabella3.2.8} and \ref{tabella3.2.9}, bug introduction is
mainly due to updating and adding CUs. This is valid for each metric
considered.

\section{Conclusion}\label{Conclusion}
A statistical description of large software systems as directed
graphs can provide much additional information on the system
features with respect to more traditional approaches, from the
software engineering perspective. Adopting
 a graph as a model for the software system, we used the compilation units as
the basic software module in order to build a software graph, and
redefined the CK
 suite of metrics to cope with CUs. These metrics were then used to investigate,
 with a statistical analysis, how and where bugs were introduced into two big, OO
  software projects like Eclipse and Netbeans. We wrote two different parsers to
  analyze the CVS log file and the issue tracker repositories in order to
   automatically associate bugs and CUs. In this paper, we introduced the concept
    of compilation unit graph, and of OO metrics related to compilation units,
    with the purpose of analyzing software projects managed using a configuration
     management system and a corresponding bug tracking system.\\
The picture of the software system as a graph allowed us to detect
fat-tail
      distributions, well described by power-laws, for different features of the system, suggesting the same general underlying framework of many other complex networks.
       In particular, we found that bugs distribution among CUs, number of CUs affected by
        bugs, metrics distributions (namely LOCs, number of in-links and out-links of the
class graph, CK metrics WMC, CBO, RFC and LCOM), all exhibit power-laws fat-tails.\\
Inside this framework it is possible to identify strong correlations
among bugs and those metrics
 related to the number of external dependencies which, in the graph representation, are easily described as directed links.
  All these findings together indicate a possible strategy to optimize resources and efforts in software engineering for
   finding, forecasting, and fixing software defects. Once the software graph reveals the fat-tail in the relationships
   between bug and CUs, one may identify which parts of the software are the most fault-prone and focus fixing efforts
   on them. Following \cite{Louridas}, if one ranks CUs according to these power-laws, the review of a small fraction among
the highest ranked may have an exponential impact on the overall amount of software defects detectable and fixable.\\
Our analysis goes one step further, examining software evolution
across many releases. This study identifies how metric evolution is
related to bug introduction. The change of some particular metrics
may result in a higher probability
 of introducing a bug. In particular, we identify different families of CUs, related to CK metrics, showing different
 robustness with respect to bug affection. Our categorization into families is related to the software evolution and it
  is useful to investigate correlations among bugs and software metrics changes. This classification may be particularly
   useful to software engineers in order to decide onto which parts of a big software project it is better to concentrate
    efforts and resources.

\end{document}